\documentclass[aps,prd,showpacs,nofootinbib,floats,floatfix,preprintnumbers,groupedaddress,twocolumn]{revtex4}

\usepackage{bm}
\usepackage{latexsym}
\usepackage{dcolumn}
\usepackage{amsmath,amsfonts,amssymb}
\usepackage{graphicx,epsfig}
\usepackage{amsmath}
\usepackage{fancyhdr}
\usepackage{hyperref}
\usepackage{graphicx,epstopdf}

\usepackage{mathtools}
\DeclarePairedDelimiter\bra{\langle}{\rvert}
\DeclarePairedDelimiter\ket{\lvert}{\rangle}
\DeclarePairedDelimiterX\braket[2]{\langle}{\rangle}{#1 \delimsize\vert #2}

\hypersetup{
	colorlinks   = true, 
	urlcolor     = blue, 
	linkcolor    = blue, 
	citecolor   = red 
}

\begin{document}
\title{Detector response along null geodesics in black hole spacetimes and in a  Friedmann-Lemaitre-Robertson-Walker Universe}
\author{Kushal Chakraborty\footnote{\color{blue}kushalstan@iitg.ac.in; kushalchakraborty5@gmail.com}}
\author{Bibhas Ranjan Majhi\footnote {\color{blue}bibhas.majhi@iitg.ac.in}}

\affiliation{Department of Physics, Indian Institute of Technology Guwahati, Guwahati 781039, Assam, India}

\date{\today}

\begin{abstract}
We study the detector's response when moving along an ingoing null geodesic. The backgrounds are chosen to be black hole spacetimes ($(1+1)$ dimensional Schwarzschild metric and near horizon effective metric for any stationary black hole in arbitrary dimensions) as well as Friedmann-Lemaitre-Robertson-Walker (FLRW) Universe. For black holes the trajectories are defined in Schwarzschild coordinates and the field modes are corresponding to {\it Boulware vacuum}.  Whereas for FLRW case, the detector is moving along the path defined in original {\it cosmic time} and the field modes are related to {\it conformal vacuum}. The analysis is done for three stages (de-Sitter, radiation dominated and matter dominated) of the Universe. We find that, although the detector is freely falling, it registered particles in the above mentioned respective vacuums. We confirm this by different approaches. The detection probability distributions, in all situations, are thermal in nature. 
\end{abstract}


\maketitle

\section{Introduction}
It is now well known fact that black holes radiate \cite{Hawking:1974rv}. One noticeable observation in this context is that radiation from the horizon is an observer dependent fact. For instance, Hawking radiation observed by a static observer at infinity; whereas a freely falling frame does not perceive any thermality. This fact is well supported by the Unruh effect \cite{Unruh:1976db} where an uniformly accelerated observer (Rindler frame) \cite{Rindler:1960zz} sees the Minkowski vaccum as thermal state. All these resemblance between these two facts is well compatible with the {\it equivalence principle}. Several recent investigations has been done related to the nature of the observed particles in this thermal bath \cite{Adhikari:2017gyb,Chowdhury:2017ifj,Das:2019aii,Chowdhury:2019set}. A very recent investigation reveled that these particles can exhibit Brownian like motion in this thermal bath and satisfy the well known equilibrium fluctuation-dissipation theorem \cite{Adhikari:2017gyb,Das:2019aii}.   

Recently, Scully et al \cite{Scully:2017utk} explicitly showed an radially freely in-falling atomic detector will see {\it Boulware vacuum} as thermal bath. Although apparently it seems that this process signifies the breakdown of equivalence principle as there is no gravitational acceleration, but this is indeed not. Actually, the relative acceleration between the Boulware field modes and the detector plays the trick. This fact is well supported by some earlier sporadic attempts \cite{Frolov,Ahmadzadegan:2013iua}. It may be noted that the work in \cite{Scully:2017utk} is confined to Schwarzschild black hole and the path of the detector were chosen to be {\it timelike}.

In this present paper, we investigate different cases within the setup of \cite{Scully:2017utk}. First of all in every situation, the ingoing path is taken to be {\it null like}. Considering the detector to be moving along this path, we investigated its response for the black hole as well as Friedmann-Lemaitre-Robertson-Walker (FLRW) backgrounds. For the black hole metric, the modes are chosen to be Boulware ones and detector moves radially in null path. Here we start with $(1+1)$ dimensional Schwarzschild black hole and then the same has been extended to the near horizon effective metric of any arbitrary dimensional stationary black hole. This basically generalizes the investigation. For the later case, the detector is allowed to move with the near horizon region whereas in the Schwarzschild one it moves from infinity to the horizon. 

We also discuss the detector response in three stages (de-Sitter (dS) era, radiation dominated era and matter dominated era) of the FLRW Universe. The field mode under investigation is the {\it conformal mode} which is obtained by the solution of the massless Klein-Gordon (KG) equation in conformal time coordinate. The corresponding vacuum is called as the {\it conformal vacuum}. We find that the detector, which is following the ingoing null path in original {\it cosmic time} coordinate, perceives particles in the conformal vacuum. It is interesting to mention that in this analysis for dS universe, the excitation probability depends only on detector's frequency; whereas others are function of both mode frequency as well as that of detector. For later ones, probability decreases with the increase of frequency of the detector. 

The organization of the paper is as follows. In the next section, we briefly setup main idea and formula of the detector's response function. This quantity is first calculated and analyzed for $(1+1)$ dimensional Schwarzschild black hole in Section \ref{Null} by deriving the ingoing radial null trajectories of the detector. Sections \ref{General} and \ref{FRW} discuss the same for near horizon effective metric for any stationary arbitrary dimensional black hole and $(1+1)$ dimensional FLRW Universe, respectively. In section \ref{App3} we analysis the $(1+3)$ dimensional FLRW situation for a massless, conformally coupled scalar field. Finally, we conclude in Section \ref{Conclusions}. For cross verification, the particle production procedure, for the present cases, has also been analyzed in other two different approaches. This is presented in two appendices (Appendix \ref{App1} and Appendix \ref{App2}), which are added at the end.    
\section{\label{Setup}Set up: Atomic detector}
Now let us consider a two-level (say $a$ is the excited level and $b$ is the ground state) atom is moving freely along a particular geodesic in our spacetime backgrounds (black hole and FLRW Universe). Let us consider massless scalar field $\Phi$ under this background whose modes are denoted by $u_\nu$ with frequency $\nu$. The modes of the atomic detector are labeled by $\psi_\omega$, where $\omega$ is its characteristic frequency. Then the interaction Hamiltonian between the field and the atomic detector is given by \cite{Scully:2017utk} (see \cite{Book1}, for the details of the detector response),
\begin{equation}
\hat{V}(\lambda)=g\left[(\hat{a}_{\nu}u_\nu+h.c.)(\hat{\sigma}_{\omega}\psi_\omega+h.c.)\right]~.
\label{2.01}
\end{equation}
In the above, the operator $\hat{a}_{\nu}$ is the photon annihilation operator and $\hat{\sigma}_{\omega}$ is the atomic detector lowering operator. Here $g$ is the coupling constant, determines the strength of interaction, whereas $h.c.$ signifies the hermitian conjugate. $\lambda$ is the detector's clock time.

Initially, the atomic detector is in the ground state and there are no photons for the field mode frequency $\nu$; i.e. the field is in the Boulware vacuum (which gives an empty spacetime for a stationary observers) for the black hole and conformal vacuum for the FLRW Universe. We denote the Boulware vacuum or the conformal vacuum and the one particle state of the field $\Phi$ as $\ket{0}$ and $\ket{1_{\nu}}$, respectively. Whereas the ground and the excited states of the atomic detector are labeled by $\ket{b}$ and $\ket{a}$. Now the transition ($\ket{b}\rightarrow \ket{a}$) amplitude of the detector for the detection of one scalar particle state, at the first order perturbation theory is given by
\begin{equation}
\Gamma = -i\int_{\lambda_i}^{\lambda_f} d\lambda \bra{1_{\nu},a}\hat{V}(\lambda)\ket{0,b}~,
\label{2.02}
\end{equation}
where $\lambda$ is the detector's clock time. For a massive detector, it is the proper time while for massless one it is the {\it affine parameter} which defines the four-momentum $p^a$ of the detector in such  a way that it satisfies $p^a\nabla_ap^b=0$ with $p^a=dx^a/d\lambda$.  
Therefore, the probability of excitation of the atomic detector, at this order, for the interaction Hamiltonian (\ref{2.01}) turns out to be
\begin{eqnarray}
P_{\uparrow} & =&\Big|\int_{\lambda_i}^{\lambda_f} d\lambda \bra{1_{\nu},a}\hat{V}(\lambda) \ket{0,b} \Big|^{2}
\nonumber \\
 & =&g^{2}\Big|\int_{\lambda_i}^{\lambda_f} d\lambda u_\nu^*(\lambda)\psi_\omega^*(\lambda)\Big|^{2}~.
 \label{2.03}
 \end{eqnarray}
This is our working formula and will be re-expressed in different forms according to the situations.
 
For the black holes, here we are interested only on the radial trajectories of the atomic detector which will approach towards horizon. Therefore the variable $\lambda$ of the above equation has be expressed in terms of radial coordinate $r$ (say) and the integration limit of $r$ has to be from initial value of $r$ (say $r_i$) to horizon $r_H$. Under this circumstances, we re-express (\ref{2.03}) as 
\begin{equation}
 P_{\uparrow} =g^{2}\Big|\int_{r_i}^{r_{H}}dr\Big(\frac{d\lambda}{dr}\Big)u_\nu^*(r)\psi_\omega^*(r)\Big|^{2}~.
 \label{2.04}
\end{equation}
This we will explicitly evaluate by considering the {\it ingoing} radial motion of the atomic detector. For a particular background metric one has to first find the expressions for the modes $u_\nu$ and $\psi_\omega$. Then using the solutions of the equations of motion for the detector trajectory express everything in terms of radial coordinate. Two types path are allowed: timelike and null like. Detector response for the timelike path has been extensively studied in \cite{Scully:2017utk} for a Schwarzschild black hole. Here we shall concentrate null like paths in a more general background. 

In the later discussion, we shall start our analysis with the ($1+1$) dimensional Schwarzschild background where the detector is moving from infinity to the horizon. Then the same will be extended to more general background with arbitrary dimensions by confining the motion of the detector very close to the horizon. Also the three stages of the FLRW Universe will be discussed. For these, we, in the next section will find the null radial trajectories for the detector.  

\section{\label{Null}Null radial trajectories for the detector: $(1+1)$ dimensional Schwarzschild black hole}
The Schwarzschild metric in $(1+1)$ dimensional space-time in terms of Schwarzschild coordinates $(t_s,r)$ is given by
\begin{equation}
ds^{2}=-f(r)dt_{s}^{2}+\frac{dr^{2}}{f(r)}~,
\label{2.05}
\end{equation}
where $f(r)=(1-\frac{r_{H}}{r})$ with the horizon is located at $r_{H}=2M$. Here $M$ is the mass of the black hole.
To remove the coordinate singularity at $r=r_H$, consider the Painleve coordinate transformation:
\begin{equation}
dt_{s}= dt_{p}-\frac{\sqrt{1-f(r)}}{f(r)}dr~.
\label{2.06}
\end{equation}
Under this transformation the metric will take the form,
\begin{equation}
ds^{2}=-f(r)dt_{p}^{2}+2\sqrt{1-f(r)}dt_{p}dr+dr^{2}~.
\label{2.07}
\end{equation}
As there is no explicit $t_p$ dependence in the metric coefficients, there will be a timelike Killing vector $\chi^{a}=(1,0,0,0)$. Hence the energy of the particle moving in this background is given by $E=-\chi^{a}p_{a}=-p_{t_{p}}$, where $p_{t_p}$ is the covariant time component of the momentum
$p_{a}=(p_{t_{p}},p_{r})$.

Now in order to find the null trajectory, we shall start with the dispersion relation $g^{ab}p_ap_b=0$ for a massless particle, where the contravariant component of momentum is defined as $p^a=dx^a/d\lambda$ with $\lambda$ is identified as the affine parameter such that $p^a\nabla_ap^b=0$ is satisfied. Expanding the dispersion relation for the metric (\ref{2.07}) and replacing $p_{t_p}=-E$ we obtain
\begin{equation}
E^{2}-f(r)p_{r}^{2}+2Ep_{r}\sqrt{1-f(r)}=0~.
\label{2.08}
\end{equation}
Solution of the above for $E$ yields
\begin{equation}
E=-\sqrt{1-f(r)}p_{r}\pm p_{r}~.
\label{2.09}
\end{equation}
Here the positive sign stands for the outgoing trajectory while the negative one refers to ingoing trajectory. Concentrating only on the ingoing one and using the Hamilton's equation of motion, one finds the radial equation as
\begin{equation}
\dot{r}=\frac{dr}{d\lambda}=\frac{\partial E}{\partial P_{r}}=-\sqrt{1-f(r)} - 1~.
\label{2.10}
\end{equation}
Also we have $ds^{2}=0$ for a null path. This yields, for ingoing path, $(dt_p/dr)$ as
\begin{equation}
\frac{dt_{p}}{dr}=\frac{\sqrt{1-f(r)}-1}{f(r)}~.
\label{2.11}
\end{equation}
These equations (\ref{2.10}) and (\ref{2.11}) will help us to express the affine parameter $\lambda$ and the coordinate time $t_p$ in terms of radial coordinate $r$.

Substituting the expression for $f(r) = 1-r_H/r$ in (\ref{2.10}) and (\ref{2.11}) and then integrating 
we find
\begin{eqnarray}
\lambda  &=&-r+2\sqrt{r_Hr}-2r_{H}\ln\left[\frac{\sqrt{r}}{\sqrt{r_{H}}}+1\right]~;
\label{2.12}
\\
t_{p}&=&-r+2\sqrt{r_Hr}-2r_{H}\ln\left[\frac{\sqrt{r}}{\sqrt{r_{H}}}+1\right]~,
\label{2.13}
\end{eqnarray}
up to some irrelevant integration constant. The above implies that that the Painleave time $t_p$ can be identified as the affine parameter for the null path.
For our main purpose, we shall notice that the relation between the Schwarzschild time $t_s$ and radial coordinate $r$ is needed. This can be obtained by finding the relation among $t_s$ and $t_p$. Integrating (\ref{2.06}) for the present value of $f(r)$ one obtains
\begin{equation}
t_{p}=t_{s}+2\sqrt{r_Hr}+r_{H}\ln\,\left(\frac{\sqrt{\frac{r}{r_{H}}}-1}{\sqrt{\frac{r}{r_{H}}}+1}\right)~.
\label{2.14}
\end{equation}
Substituting this in (\ref{2.13}), we find the required expression:
\begin{equation}
t_{s}  =-r-r_{H}\ln\,(\frac{r}{r_{H}}-1)+{\textrm{constant}}~.
\label{2.15}
\end{equation}
The equations (\ref{2.12}) and (\ref{2.15}) represent the trajectory of the null-like detector incoming radially towards the black hole. We shall use them in the explicit evaluation of (\ref{2.04}).
\section{Evaluation Of Detector Response Function}
In this section we try to study detector response function while using results developed in the previous sections. To do so we first need to find our $u_\nu$ and $\psi_\omega$.

\subsection{Detector and scalar field mode functions}
 The positive frequency mode corresponding to the detector is 
 \begin{equation}
 \psi_\omega=e^{-i\omega\lambda}~.
 \label{2.16}
 \end{equation}
 The positive frequency mode for massless scalar field can be obtained by solving the Klein-Gordon (KG) equation $\Box\Phi=0$ under the background (\ref{2.05}). This is easily solved in the Regge-Wheeler coordinates,
\begin{equation}
r^{*}(r)=r+r_{H}\ln\left(\frac{r}{r_{H}}-1\right)~.
\label{2.17}
\end{equation}
The interval (\ref{2.05}) in this new coordinate looks like
\begin{equation}
ds^{2} =f(r^*)[-dt_{s}^{2}+d{r^*}^{2}]~,
\label{2.18} 
\end{equation}
and the KG equation reduces to the following form:
\begin{equation}
\left[\frac{\partial^{2}}{\partial t_{s}^{2}}-\frac{\partial^{2}}{\partial {r^*}^{2}}\right]\Phi=0~.
\label{2.19}
\end{equation}
It leads to two solutions $e^{- i\nu(t_s\pm r^*)}$, where positive sign corresponds to ingoing and negative sign refers to outgoing modes. Since only the outgoing modes can be detected by the detector (as ingoing modes enter into the black hole and is trapped in this region), we consider the mode solution of scalar field as 
\begin{equation}
u_\nu=e^{-i\nu(t_s-r^{*})}~.
\label{2.20}
\end{equation}
Note that it is also the positive frequency mode. This is our Boulware field mode.

\subsection{Evaluation of probability of excitation}
In order to evaluate (\ref{2.04}), we have to substitute (\ref{2.10}) with $f(r)=1-r_H/r$ and the expressions for modes from (\ref{2.16}) and (\ref{2.20}). Expressing everything in terms of radial coordinate by using our path, given by equations (\ref{2.12}) and (\ref{2.15}), one obtains
\begin{eqnarray}
P_{\uparrow} &&=g^{2}\Big|\int_{\infty}^{r_{H}}dr\Big(-\frac{\sqrt{\frac{r}{r_{H}}}}{1+\sqrt{\frac{r}{r_{H}}}}\Big)e^{i\nu\Big[-2r-2r_{H}\ln\,(\frac{r}{r_{H}}-1)\Big]}
\nonumber
\\
&&\times e^{i\omega\Big[-r+2\sqrt{r_Hr}-2r_{H}\ln\left(\frac{\sqrt{r}}{\sqrt{r_{H}}}+1\right)\Big]}\Big|^{2}~,
\label{2.21}
\end{eqnarray}
where we have chosen the lower limit $r_i$ of the integration as infinity as the detector starts from very far from the horizon. 

The above integration can not be evaluated analytically. Of course this can be analytically tackled under a particular approximation. To get a feel of this response function (\ref{2.21}), we first evaluate it under very large $\omega$ region. The same has also been done in \cite{Scully:2017utk}, where the path of the detector was taken as timelike. Later we shall analyse the above numerically without any approximation. 

\subsubsection{Analytical Approach: large $\omega$ limit}
To evaluate  equation (\ref{2.21}) analytically, we first change $r$ variable to a suitable one: $\sqrt{r/r_H}=z$. Then (\ref{2.21}) reduces to the following form: 
\begin{eqnarray}
P_{\uparrow} &=& g^{2}r_{H}^{2}\Big|\int_{\infty}^{1}\left(\frac{2z^2dz}{1+z}\right)e^{i\nu[-2r_Hz^2-2r_H\ln(z^2-1)]}
\nonumber
\\
&\times& e^{i\omega[-zr_{H}(z-2)-2r_{H}\ln(z+1)]}\Big|^{2}~.
\label{2.22}
\end{eqnarray}
This integration can not be performed at this stage. But we shall see that an approximate expression can be obtained for large $\omega$ (more precisely, $r_H\omega\gg1$). The same has also been adopted in \cite{Scully:2017utk} for timelike path. To implement this approximation, we do a variable substitution $x=r_H\omega(z-1)$.
Then (\ref{2.22}) turns out to be
\begin{eqnarray}
P_{\uparrow} &=& g^2r_{H}^{2}\Big|\int_{0}^{\infty}\frac{dx}{r_H\omega} \Big(1+\frac{x}{r_H\omega}\Big)^2\Big(1+\frac{x}{2r_H\omega}\Big)^{-1}
\nonumber
\\
&\times&e^{i\nu[-2r_{H}(1+\frac{x}{r_H\omega})^2-2r_{H}\ln(\frac{x}{r_H\omega})-2r_{H}\ln(1+\frac{x}{2r_H\omega})]}
\nonumber
\\
&\times& e^{i\omega[r_{H}(1-\frac{x^2}{r_H^2\omega^2})-2r_{H}\ln(1+\frac{x}{2r_H\omega})]}\Big|^{2}~.
\label{2.23}
\end{eqnarray}
In the limit $r_H\omega\gg 1$, keeping only upto first order in $x/(r_H\omega)$ terms in the integrant one obtains
\begin{equation}
P_{\uparrow} \simeq \frac{g^2r_{H}^{2}}{r_H^2\omega^2}\Big|\int_0^{\infty}dx x^{-2i\nu r_H} e^{-i(\frac{5\nu}{\omega}+1)x}\Big|^2~.
\label{2.24}
\end{equation}
It is now straightforward to compute by using the standard integration result (see page $604$ of \cite{Paddybook} for details):
\begin{equation}
\int_{0}^{\infty}x^{s-1}e^{-bx}dx=\exp(-s\ln b)\Gamma(s)~,
\label{2.25}
\end{equation}
where Re $s>0$ and Re $b>0$. Using this one finds
\begin{equation}
P_{\uparrow} \simeq \frac{4\pi g^2r_H\nu}{\omega^2(1+\frac{5\nu}{\omega})^2}\frac{1}{e^{4\pi r_H\nu}-1}~.
 \label{2.26}
\end{equation}
The evaluation of the integration in (\ref{2.24}) is as follows. If one compares this with (\ref{2.25}), one finds $s=1+2i\nu r_H$ and $b=i(5\nu/\omega + 1)$. The value of $b$ makes the integration divergent. To ensure the convergence consider the value of  $b$ as
\begin{equation}
\,\,\,\,\ b=i(5\nu/\omega + 1)+\epsilon~,
\label{RF1}
\end{equation}
with the limit $\epsilon\rightarrow 0^{+}$. Then one has
\begin{eqnarray}
&&\ln b =\lim_{\epsilon\rightarrow 0^+}\ln[i(\frac{5\nu}{\omega}+1)+\epsilon]
\nonumber
\\
&&=\ln|\frac{5\nu}{\omega}+1|+\frac{i\pi}{2}\textrm{sign}(\frac{5\nu}{\omega}+1)~,
\label{RF2}
\end{eqnarray}
where ``sign'' is the sign function. Using this we obtain
\begin{eqnarray}
&&\int_0^{\infty}dx x^{-2i\nu r_H} e^{-i(\frac{5\nu}{\omega}+1)x}
\nonumber
\\
&&=(1+\frac{5\nu}{\omega})^{-(1+2i\nu r_H)}e^{-\frac{i\pi}{2}}e^{-\pi\nu r_H}\Gamma(1+2i\nu r_H)~.
\label{RF3}
\end{eqnarray}
Substitution of this in (\ref{2.24}) yields (\ref{2.26}). The same prescription will also be followed later.

So we see that the probability of excitation of the detector is non-zero and hence there must be particle production in the Boulware vacuum with respect to the radially infalling detector. Hence Boulware vacuum appears to be thermal with temperature identical to Hawking temperature $T = 1/(4\pi r_H)$. It must be  emphasized that there is a crucial difference between this one with that for the standard Unruh-De-Witt detector. The exponential factor here depends on the field frequency $\nu$; whereas the same for Unruh-De-Witt case depends on the energy gap of the detector levels. 
Now, below we want to try a more holistic approach, without any approximation. For that we shall take the help of the numerical technique.

\subsubsection{Numerical approach}
The above approximate calculation suggests that there is a thermal bath with respect to the radially infalling observer, and this is valid for large values of detector's frequency. Now to have better understanding of the expression (\ref{2.21}) in all values of $\omega$, here we shall examine this numerically. First and foremost to tackle this
numerically we need to make all the variables dimensionless.
For that we choose the following substitutions in equation (\ref{2.22}):
\begin{equation}
 r_H\omega = \omega'; \,\,\,\ r_H\nu = \nu'~.
\label{2.27}
\end{equation}
Then (\ref{2.22}) reduces to the following form: 
\begin{eqnarray}
\frac{P_{\uparrow}}{g^2r_{H}^{2}}&&=\Big|\int_{\infty}^{1}\left(\frac{2z^{2}dz}{1+z}\right)e^{i\nu'[-2z^{2}-2\ln(z-1)-2\ln(z+1)]}
\nonumber
\\
&&\times e^{i\omega'[-z^{2}+2z-2\ln\left(z+1\right)]}\Big|^{2}~.
\label{2.28}
\end{eqnarray}
Next consider a variable substitution of the form $z-1=x$; Finally to obtain a more convenient form further substitute $x(x+2)=y$. This yields
\begin{eqnarray}
P'_{\uparrow} &&=\Big|\int_{0}^{\infty}dy\left(\frac{\sqrt{1+y}}{\sqrt{1+y}+1}\right)e^{-(2i\nu'+\epsilon) y}y^{-2i\nu'}
\nonumber
\\
&&\times e^{-i\omega'[\sqrt{1+y}(\sqrt{1+y}-2)]}(\sqrt{1+y}+1)^{-2i\omega'}\Big|^{2}~,
\label{2.29}
\end{eqnarray}
where we denoted $P'_{\uparrow}=P_{\uparrow}/g^2r_{H}^{2}$. In the above we inserted a $\epsilon$ parameter which is $\epsilon\rightarrow 0^+$ to make the integrant convergent within the limits of integration.

\begin{figure}[h!]
\begin{centering}
\includegraphics[scale=0.40]{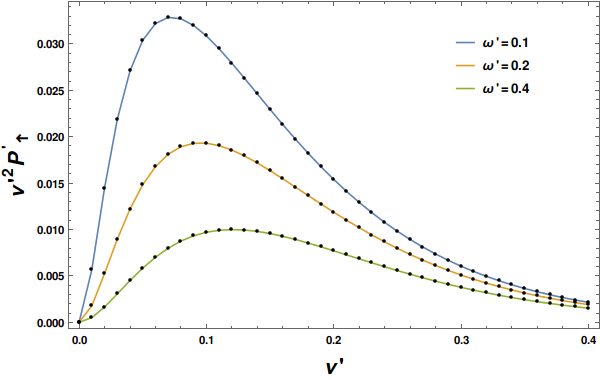}
\end{centering}
\caption{Plot of $\nu'^2P'_{\uparrow}$ Vs $\nu'$ for different values of $\omega'$. Here we choose the small parameter $\epsilon=0.01$.}
\label{Fig1}
\end{figure}

Now with the help of {\it Mathematica} package we numerically integrate the above expression for different constant values of $\omega'$ and then plot $\nu'^2 P'_{\uparrow}$ as a function of $\nu'$. This is represented in Fig. \ref{Fig1}. 
As expected the plots show Planck distribution type nature. So the detector must register particles in the Boulware vacuum. As we keep increasing $\omega'$, the peak keeps going lower and the curve covers
less area. This suggests that at higher $\omega'$ value the chances
of particle detection is lower.   Also the the approximate analytical
approach that we took to deal the problem suggested Planck's type distribution.
Hence this numerical solution plot also features the same characteristics.
From this plot one thing is evident that for higher value of $\omega'$ the transition
probability decreases substantially.

\section{\label{General}A general black hole in arbitrary dimensions: near horizon analysis}
The earlier discussion was done for the Schwarzschild black hole in $(1+1)$ dimensional background. Now we want to extend the same discussion for other black holes in arbitrary dimensions. It must be mentioned that simple extension to this situation is very complicated  and very difficult to analyse. But we can simplify the situation by considering the near horizon geometry of the black holes. As we are interested to radial motion of the atomic detector and want to examine if this can detect particles in the Boulware vacuum, the above approximation is sufficient. There will not be any loss of generality for the present discussion. {\it Only we have to move the detector within our region of approximation} (not like earlier one from infinity to horizon). So the detector will fall radially from a near horizon radial point ($r$, say) such that $(r/r_H-1)<<1$  to the horizon $r_H$.

It is well known that the black hole spacetimes, near the horizon, is effectively ($1+1$) dimensions \cite{Carlip:1998wz,Robinson:2005pd,Iso:2006wa} (also see \cite{Majhi:2011yi,Banerjee:2010ye}, for more discussions and references). The idea is the following. If one expands the massive Klein-Gordon (KG) action (taken for simplicity) for a general black hole background, in the near horizon limit the action reduces to the form which is similar to massless KG equation under the effective background (\ref{2.05}). Only the explicit expression of $f(r)$   is different for different black holes. For example for Kerr-Newman one it is given by $f(r) = (r^2-2Mr+a^2+e^2)/(r^2+a^2)$ where $M, e, a$ are the mass, charge and rotation parameters of the black hole, respectively \cite{Umetsu:2009ra}. The near horizon form (\ref{2.05}) also includes anti-de-Sitter (AdS) black holes. Now since we are interested to near horizon region, the metric coefficient $f(r)$ can be taken as the first leading term of the Taylor series expansion of it around $r=r_H$:
\begin{equation}
f(r) \simeq 2\kappa(r-r_H)~.
\label{2.30}
\end{equation}
Here $\kappa=f'(r_H)/2$ is the surface gravity which includes the explicit information about the particular black hole. We shall see that the explicit expression of $\kappa$ will not be needed to achieve our main goal. 

The null paths, in this case, again given by (\ref{2.10}) and (\ref{2.11}) with $f(r)$ is now identified as (\ref{2.30}). Therefore, $\dot{r}$ turns out to be
\begin{equation}
\dot{r} = \frac{dr}{d\lambda} = -1-\sqrt{1-2\kappa(r-r_H)} \simeq \kappa(r-r_H)-2~.
\label{2.31}
\end{equation}
Integrating this, upto linear order in $(r/r_H - 1)$, we obtain
\begin{equation}
\lambda \simeq -\frac{r}{2}~.
\label{2.32}
\end{equation}
Similarly, Eq. (\ref{2.11}), upto this order, yields $t_p\simeq -r/2$. Hence using (\ref{2.06}) the Schwarzschild time coordinate $t_s$ turns out to be
\begin{equation}
t_s \simeq t_p + \frac{r}{2}-\frac{1}{2\kappa}\ln\Big(\frac{r}{r_H}-1\Big)~.
\label{2.33}
\end{equation} 
In this case, the tortoise coordinate $r^*$ in terms of radial coordinate turns out to be
\begin{equation}
r^*=\frac{1}{2\kappa}\ln\Big(\frac{r}{r_H}-1\Big)~.
\label{2.34}
\end{equation} 
Here again the mode functions are of the form (\ref{2.16}) and (\ref{2.20}). Substituting all these in the general formula (\ref{2.03}), we find
\begin{eqnarray}
P_{\uparrow} &=& g^2\Big| \int_{r_i}^{r_H} \frac{dr}{1+\sqrt{1-2\kappa(r-r_H)}}
\nonumber
\\
&\times&\Big(\frac{r}{r_H}-1\Big)^{-i(\nu/\kappa)} e^{-i(\omega r/2)}\Big|^2~.
\label{2.35}
\end{eqnarray}
In the above $r_i$ has to chosen such that it's value satisfies our near horizon approximation; i.e. $(r/r_H - 1)<<1$.

\begin{figure}[h!]
\begin{centering}
\includegraphics[scale=0.40]{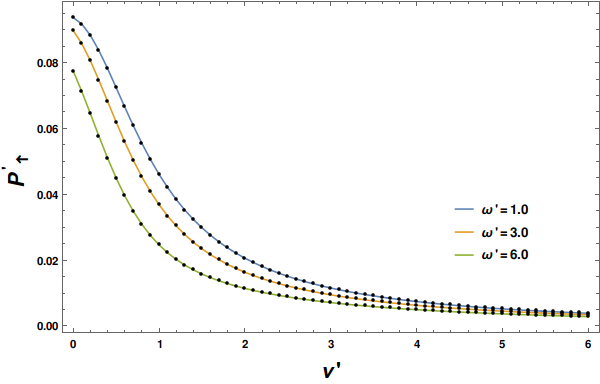}
\end{centering}
\caption{Plot of $P'_{\uparrow}$ Vs $\nu'$ for different values of $\omega'$. Here we choose the small parameter $\epsilon=0.001$, $x_i=0.5$ and $\kappa'=1$.}
\label{Fig2}
\end{figure}

 To make the above expression in a convenient form, let us first change the variable: $(r/r_H) - 1=x$. Then (\ref{2.35}) reduces to 
\begin{eqnarray}
P_{\uparrow} &=& g^2\Big| \int_{0}^{x_i} \frac{r_H dx}{1+\sqrt{1-2\kappa r_H x}}
\nonumber
\\
&\times&(x)^{-i(\nu/\kappa)} e^{-i(\omega r_H/2)(1+x)}\Big|^2~.
\label{2.36}
\end{eqnarray}
In terms of dimensionless parameters $\omega'=r_H\omega$, $\nu'=\nu/\kappa$ and $\kappa' = \kappa r_H$, we find the probability of excitation as 
\begin{equation}
P'_{\uparrow} = \Big| \int_{0}^{x_i} \frac{dx}{1+\sqrt{1-2\kappa' x}}
(x)^{-i\nu'} e^{-(\frac{i\omega'}{2}+\epsilon)(1+x)}\Big|^2~.
\label{2.37}
\end{equation}
where $P'_{\uparrow} = P_{\uparrow}/(g^2r_H^2)$. Now we numerically plot the above in Fig. \ref{Fig2}.
This shows that the detector register particle in Boulware vacuum and, like earlier, probability decreases with the increase of $\omega'$.

Note that in this case we did not plot $\nu'^2P'_{\uparrow}$ Vs $\nu'$; instead we plotted $P'_{\uparrow}$ Vs $\nu'$. The reason is the following. In $\nu'^2 P'_{\uparrow}$, $\nu'^2$ is a increasing function while $P'_{\uparrow}(\nu')$ is a decreasing one.  Normally for Planck case, in the range of low values of frequency, $\nu'^2$ dominates which effectively gives the increasing behaviour and after certain value of $\nu'$, $P'$ dominates and gives decreasing behaviour of the composite function. Therefore effectively we obtain the Planck type plot. But for the present situation, in our numerical calculation we observed that $\nu'^2P'_{\uparrow}$ is always an increasing function. This is because in the large values of $\nu',$ the numerical values of $\nu'^2$ term are always so large compared to that of $P'_{\uparrow}$ such that their multiplication is always increases and hence we never get the decreasing behaviour like Planck plot, although the Fig. \ref{Fig2} shows $P'_{\uparrow}$ always a decreasing function of $\nu'$. Having this limitation (which is just a numerical difficulty) and since our main aim is to see if the probability is non-trivial, we just show the nature of $P'_{\uparrow}$ as a function of $\nu'$ which is sufficient to argue that  the probability of transition is non-vanishing. Interestingly, the nature is similar to Bose-Einstein distribution and this tells that the vacuum appears to be thermal with respect to our infalling observer. In later cases as we shall see, fortunately such difficulty does not arise.

Before concluding this section, let us comment on the choice of the vacuum (here it is Boulware) when the detector is in the near horizon region. Boulware vacuum is defined with respect to a static observer in the $(t,r^*)$ coordinates in which the scalar modes are given by (\ref{2.20}), called as Boulware modes.  This vacuum, asymptotically at $r\rightarrow\infty$, is Minkowski vacuum. So a static observer at $r\rightarrow\infty$, where the spacetime is higher dimensional, defines this vacuum as Minkowski. Now we considered another observer with an atomic detector is freely falling towards the horizon from a point within the near horizon regime. This detector is interacting with the Boulware modes. As we said, this observer will see this effectively $(1+1)$ dimensional spacetime when it is constrained to move within this limit. For this observer, what we found is, the Boulware mode appears to be thermal. So here we have two observers: one is defining the Boulware vacuum and the scalar modes corresponding this are investigated by another observer who is moving within the near horizon region. This can also be interpreted as only the radial motion of the later observer with its detector has to be turned on when the observer is within this region.   
\section{\label{FRW}Null path in FRW Universe and detector response: $(1+1)$ spacetime dimensions}
Having discussions on black hole spacetimes, now we move to the metric for the homogeneous isotropic Universe.
The FRW metric in $(1+1)$ dimensions is given by
\begin{equation}
ds^2 = -dt^2 + a^2(t)dx^2~,
\label{2.38}
\end{equation}
where the scale factor in different eras of the Universe are as follows:
\begin{eqnarray}
a(t)=\left\{\begin{array}{cl}
e^{(t/\alpha_d)},&{\mbox{de-Sitter era}}\\
C_0t^{1/2},& \mbox{Radiation dominated era}\\
C_0t^{2/3}, &\mbox{Matter dominated era}~.\end{array}\right.
\label{2.39}
\end{eqnarray}
Note that in the above $C_0$ is a dimensionfull constant. As the scale factor has to be dimensionless, for radiation dominated era, it has dimension of $t^{-1/2}$ whereas the same for matter dominated era is $t^{-2/3}$. 
In conformal time $d\eta = dt/a(t)$, the metric (\ref{2.38}) takes the form as
\begin{equation}
ds^2 = a^2(\eta)(-d\eta^2 + dx^2)~,
\label{2.40}
\end{equation}
in which the solution of the massless KG equation leads to following positive frequency outgoing mode:
\begin{equation}
u_\nu = e^{-i\nu(\eta-x)}~.
\label{2.41}
\end{equation}
We call this mode as {\it conformal mode} and corresponding vacuum as {\it conformal vacuum}. Our target is to investigate this conformal vacuum from the perspective of an atomic detector which is moving along the null trajectory in $(t,x)$ coordinates.

Before going to the main discussion, let us briefly discuss the reason behind calling the above mode as conformal mode. The Lagrangian density for a scalar field $\phi$ coupled to gravity is
\begin{equation}
\mathcal{L}=\frac{\sqrt{-g}}{2}\Big[\nabla_a\phi\nabla^a\phi - (m^2+\xi R)\phi^2\Big]~,
\label{1}
\end{equation}
where $m$ is the mass of $\phi$, $\xi$ is the coupling constant and $R$ is the Ricci scalar. $\xi=0$ is the minimally coupled case while 
\begin{equation}
\xi= \frac{n-2}{4(n-1)}~,
\label{2}
\end{equation}
is known as conformally coupled case. Here $n$ is the spacetime dimensions. For the later case, if $m=0$, then the action as well as the field equation for $\phi$ is invariant under the transformations $g_{ab}(x)\rightarrow g'_{ab}(x) = \Omega^2(x)g_{ab}(x)$ and
\begin{equation}
\phi(x)\rightarrow \phi'(x) = \Omega^{(2-n)/2}(x)\phi(x)~,
\label{3}
\end{equation}  (see discussion in section $3.2$ of \cite{Book1}). Now for $g_{ab}=\eta_{ab}$ (Minkowski metric); i.e. $g'_{ab}$ is conformally Minkowski spacetime, then the $\phi'$ is solution of scalar field equation in $g'_{ab}$ while $\phi$ is the solution for the same in $\eta_{ab}$. So for the conformal coupling case, the mode solutions in $g'_{ab}$ can be obtained by knowing the same in $\eta_{ab}$ by using the relation (\ref{3}). Consequently, the Minkowski vacuum is also the vacuum for $\phi'$. Therefore the modes are called here as conformal modes and the corresponding vacuum is known as conformal vacuum (see discussion in section $3.7$ of \cite{Book1}). Now for $(1+1)$ dimensional situation $(n=2)$, both minimal coupling and conformal coupling cases coincides as $\xi=0$ (see Eq. (\ref{2})). While for $(1+3)$ dimensional (which we shall discuss in the next section) case such thing does not happen. Since for conformal coupling case the scalar modes are easily obtainable by knowing those in Minkowski spacetime, we keep our discussion to this simple situation for the moment, although it does not matter for $(1+1)$ dimensional FLRW metric. 

The {\it ingoing} null path for the metric (\ref{2.38}) is $dt = -a(t)dx$, which in terms of conformal time turns out to be
\begin{equation}
\eta = -x~.
\label{2.42}
\end{equation}
Here, the detector mode is again given by (\ref{2.16}) and one can verify that the detector's time $\lambda$ in this case is $t$. Then the probability of excitation of the detector (\ref{2.03}) takes the following form:
\begin{equation}
P_{\uparrow} = g^2\Big|\int_{t_i}^{t_f} dt~e^{2i\nu\eta+i\omega t}\Big|^2~.
\label{2.43}
\end{equation} 
Now we shall evaluate the above integration for the three stages of the Universe by substituting the respective values of conformal time $\eta$ in terms of detector time $t$.

\subsection{de-Sitter era}
\label{DS}
For de-Sitter Universe, the relation between $\eta$ and $t$ is $\eta=-(\alpha_d)e^{-(t/\alpha_d)}$. Here the limits of integration variable can be $-\infty$ to $+\infty$. Substituting this in (\ref{2.43}) and then changing the variable to $x=e^{-(t/\alpha_d)}$ we obtain
\begin{equation}
P_{\uparrow} = g^2\alpha_d^2\Big|\int_0^{\infty} dx e^{-2i\nu x\alpha_d} (x)^{-i\omega\alpha_d-1}\Big|^2~.
\label{2.44}
\end{equation} 
To evaluate this integration, as earlier, we need to again make it convergent by using $\epsilon$ prescription as discussed below Eq. (\ref{2.26}) and the finally needs to take $\epsilon\rightarrow 0^+$ limit. Here we identify $s=-i\omega\alpha_d+\epsilon$ and $b=2i\nu\alpha_d+\epsilon$ with $\epsilon>0$. $\epsilon$ has been added in order to make the integration convergent.
Using this earlier trick we find the excitation probability of the detector as
\begin{equation}
P_{\uparrow} = \frac{2\pi g^2\alpha_d}{\omega}\frac{1}{e^{2\pi\alpha_d\omega}-1}~.
\label{2.45}
\end{equation} 
This probability exhibits thermal nature with temperature $T=1/(2\pi\alpha_d)$, which is the standard value of de-Sitter horizon temperature as obtained earlier by different methods. Note that in this case, the above result does not depend on the scalar mode frequency $\nu$. 

\subsection{Radiation dominated era}
\label{R}
Here the conformal time is given by $\eta=(2/C_0)t^{(1/2)}$. The limits of integration is chosen to be $0$ to $+\infty$; negative values of $t$ is not incorporated as for these the scale factor will be complex, which is not allowed for the existence of our FRW metric (\ref{2.38}). Then (\ref{2.43}) turns out to be
\begin{equation}
P_{\uparrow} = 4g^2\Big|\int_0^\infty dx~ x e^{i\omega(x+\frac{2\nu}{C_0\omega})^2}\Big|^2~,
\label{2.46n}
\end{equation}
where we used $x=t^{1/2}$ for the change of variable. The above one will now be numerically plotted. For that we express the above in terms of the dimensionless parameters $z=C_0x$, $\omega' = \omega/C_0^2$ and $\nu'=\nu/C_0^2$:
\begin{equation}
P'_{\uparrow} = \Big|\int_0^\infty dz~ z e^{-(-i\omega'+\epsilon)(z+\frac{2\nu'}{\omega'})^2}\Big|^2~,
\label{2.46}
\end{equation}
where $P'_{\uparrow} = (C_0^4P_{\uparrow})/(4g^2)$.
Of course, the above can be evaluated analytically, but not in a convenient form. So we shall numerically analyse this. Just for completeness, we give below the analytical expression of this integration. For that we consider the variable transformation  of the form $(z+\frac{2\nu'}{\omega'})^2=m$ to find,
\begin{equation}
P'_{\uparrow}=\left|\int_{\frac{4\nu'^{2}}{\omega'^{2}}}^{\infty}\frac{dm}{2\sqrt{m}}(\sqrt{m}-\frac{2\nu'}{\omega'})e^{i\omega'm}\right|^{2}~.
\label{2.48}
\end{equation}
The definition of upper incomplete Gamma function can be used for this evaluation which is  given by, 
\begin{equation}
\exp(-s\ln b)\Gamma(s,t)=\int_{t}^{\infty}e^{-bx}x^{s-1}dt~.
\end{equation}
Hence we can write our expression (\ref{2.48}) in terms of upper incomplete Gamma function as ,
\begin{equation}
P'_{\uparrow}=\frac{1}{4}\left|\frac{i}{\omega'}\Gamma(1,\frac{4\nu'^{2}}{\omega'})-\frac{2\nu'}{\omega'\sqrt{\omega'}}e^{(\frac{i\pi}{4})}\Gamma(\frac{1}{2},\frac{4\nu'^{2}}{\omega'})\right|^2~.
\end{equation}

\begin{figure}[h!]
	\begin{centering}
		\includegraphics[scale=0.40]{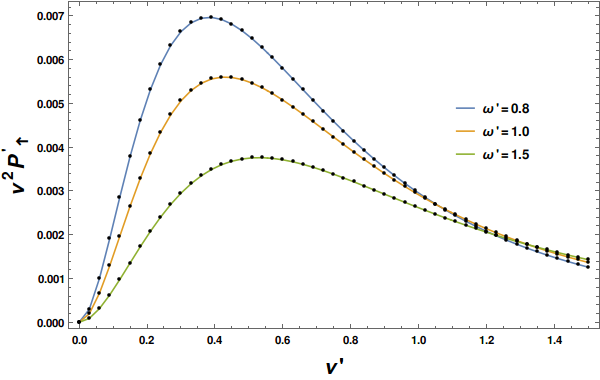}
	\end{centering}
	\caption{Plot of $\nu'^2P'_{\uparrow}$ Vs $\nu'$ for different values of $\omega'$. Here we choose the small parameter $\epsilon=0.01$.}
\label{Fig3}
\end{figure}

However to understand the nature of probability, we want to analyse numerically. For that we will use equation (\ref{2.46}). The numerical plot for $\nu'^2P'_{\uparrow}$ VS $\nu'$ is shown in Fig. \ref{Fig3}.
The feature of the curves show that the probability is indeed thermal and the detector sees the presence of particles in the conformal vacuum. Here again the particle production decreases with the increase of detector's frequency $\omega'$.

\subsection{Matter dominated era}
\label{M}
For matter dominated era, the relation between $\eta$ and $t$ is $\eta=(3/C_0)t^{1/3}$. Therefore, (\ref{2.43}) becomes
\begin{equation}
P'_{\uparrow} = \Big|\int_0^\infty dx ~ \sqrt{x} e^{-(-i\omega'+\epsilon)(x^{3/2}+\frac{6\nu' x^{1/2}}{\omega'})}\Big|^2~.
\label{2.47}
\end{equation}
Here again the limits of integration are chosen as $0$ and $\infty$ and $P'_{\uparrow}=(4C_0^3P_{\uparrow})/(9g^2)$. We also used the change of variables as $x=C_0t^{2/3}$, $\omega' = \omega/C_0^{3/2}$ and $\nu'=\nu/C_0^{3/2}$ where all these new ones are dimensionless.

\begin{figure}[h!]
	\begin{centering}
		\includegraphics[scale=0.40]{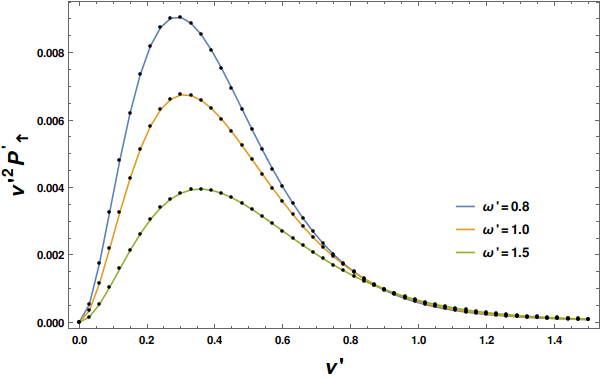}
	\end{centering}
	\caption{Plot of $\nu'^2P'_{\uparrow}$ Vs $\nu'$ for different values of $\omega'$. Here we choose the small parameter $\epsilon=0.01$.}
	\label{Fig4}
\end{figure}

We numerically plot $\nu'^2P'_{\uparrow}$ VS $\nu'$ in Fig. \ref{Fig4}.
The plot shows that the detector registers particle in the conformal vacuum. Moreover, like earlier, the detection of particle decreases with the increase of $\omega'$. 

In the next section we discuss the $(1+3)$ dimensional FLRW background case where the scalar field is considered to be conformally coupled and massless as we are interested to investigate the particle production in the conformal vacuum. Here also, as we shall see, the probabilty of transition of the detector is finite and decreases with the increasing of $\omega$.  
\section{Detector response in $(1+3)$ dimentional FLRW Universe}
\label{App3}
The FLRW metric in $(1+3)$ dimensions is
\begin{eqnarray}
ds^2&=&-dt^2+a^2(t)|d\vec{x}|^2
\nonumber
\\
&=& a^2(\eta)(-d\eta^2+|d\vec{x}|^2)~,
\label{C1}
\end{eqnarray} 
which is conformal to Minkowski metric $ds^2_M = -d\eta^2+|d\vec{x}|^2$, where $|d\vec{x}|^2=dx^2+dy^2+dz^2$. As we said earlier, to make life simple, in this background we consider conformally coupled scalar field. The minimally coupled situation in this dimensions is also very important to study. This we keep for our future study. It is well known that such a scalar field equation, in the massless situation, is conformally invariant. Moreover, the relation between the scalar fields in the Minkowski and the conformally connected backgrounds for conformally coupled massless case is 
\begin{equation}
\phi(x)=\Omega^{-1}(x)\phi_M(x)~,
\label{C2}
\end{equation}
when one finds $g_{ab}(x)=\Omega^2(x)\eta_{ab}$ and imposes the condition that the scalar equation has to be invariant. In addition, now the Minkowski vacuum is also the vacuum for $\phi(x)$, known as conformal vacuum (see, chapter $3$ of \cite{Book1} for details). Here we shall discuss the particle content in this vacuum observed by the freely moving detector along null path in the cosmic time coordinate. 

For simplicity, we consider that the detector is moving along only $x$ direction. In principle, one can consider that this detector is moving along any arbitrary direction. In that case the calculation will be more complicated. To avoid this mathematical calculation, here we assume that the detector is moving only along $x$ direction, but this will not make any loss of generality in our final goal. In this case the null ingoing path in conformal time is $x=-\eta$. The scalar mode in conformal time coordinate can be obtained by using the relation (\ref{C2}) with the identification $\Omega = a(\eta)$. The outgoing scalar mode is given by $u_\nu = a^{-1}(\eta)e^{-i(\nu\eta-\vec{p}\cdot{\vec{x}})}$ with $\nu=|\vec{p}|$. For our present choice of trajectory, it turns out to be as
\begin{equation}
u_{\nu} = a^{-1}(t)e^{-i\nu(1+\cos\theta)\eta}~,
\label{C3}
\end{equation}
where $\theta$ is the angle between detector's direction and momentum of the scalar mode. On the other hand, the detector's mode is $\psi_{\omega} = e^{-i\omega t}$. Substitution of all these in probability expression (\ref{2.03})  yields
\begin{equation}
P_{\uparrow}(\theta)=g^{2}\Big|\int_{t_i}^{t_f} dt a^{-1}(t)e^{i\nu_0\eta+i\omega t}\Big|^2~,
\label{c4}
\end{equation}
where $\nu_0=\nu(1+\cos\theta)$.

Note that the quantity inside the modulous is proportional to the transition amplitude and depends on $\theta$. We must integrate this one over the solid angle $d\Omega_0 = \sin\theta d\theta d\Phi$ with $\theta=0$ to $\theta =\pi$ and $\Phi=0$ to $\Phi=2\pi$ in order to consider all modes which have same momentum $|\vec{p}|=\nu$ but are moving in different directions. This will lead to the following expression:
\begin{eqnarray}
X =&& 2\pi\int_0^{\pi}\sin\theta d\theta\int_{t_i}^{t_f} dt a^{-1}(t)e^{i\nu_0\eta+i\omega t}
\nonumber
\\
&&=\frac{2\pi}{i\nu}\int_{t_i}^{t_f}\frac{dt}{a(t)\eta}e^{i\omega t}(e^{2i\nu\eta}-1)~.
\end{eqnarray} 
Then the transition probability of the detector comes out to be
\begin{equation}
P_{\uparrow}=\frac{4\pi^2g^{2}}{\nu^2}\Big|\int_{t_i}^{t_f}\frac{dt}{a(t)\eta}e^{i\omega t}(e^{2i\nu\eta}-1)\Big|^2\equiv\frac{4\pi^2g^{2}}{\nu^2}|Y|^2 ~.
\label{WORK}
\end{equation} 
This is our working formula for investigation different eras of the Universe.

\underline{de-Sitter era}:
Using the value of scale factor, given in Section \ref{FRW}, and the identical steps as adopted in subsection \ref{DS}, one finds 
\begin{eqnarray}
Y=-\frac{1}{\alpha_d}\int_{-\infty}^{+\infty}dt\Big[e^{-2i\nu\alpha_de^{-t/\alpha_d}+i\omega t} - e^{i\omega t}\Big]~.
\end{eqnarray}
The last term is the Dirac delta function $\delta(\omega)$ and since we are working for positive frequency modes; i.e. $\omega>0$, it will vanish.
Next, proceeding in a similar way one finds
the probability of exitation as
\begin{equation}
P_{\uparrow}= \frac{8\pi^3g^2}{\nu^2\omega\alpha_d}\frac{1}{e^{2\pi\omega\alpha_d}-1}~,
\label{C6}
\end{equation}
which reflects the thermality with the correct de-Sitter temperature. 

\begin{figure}[h!]
	\begin{centering}
		\includegraphics[scale=0.40]{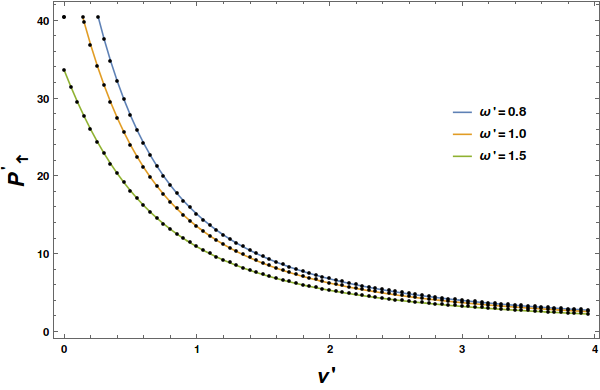}
	\end{centering}
	\caption{Plot of $P'_{\uparrow}$ Vs $\nu'$ for different values of $\omega'$. Here we choose the small parameter $\epsilon=0.01$.}
	\label{FigC1}
\end{figure}

\underline{Radiation dominated era}: Proceeding similar to subsection \ref{R} we find
\begin{equation}
P'_{\uparrow} = \frac{1}{\nu'^2}\Big|\int_0^{\infty} \frac{dx}{x}  e^{-(-i\omega'+\epsilon)x}(e^{4i\nu'x^{1/2}}-1)\Big|^2~,
\label{C7}
\end{equation}
where the following notations are introduced:
\begin{eqnarray}
P'_{\uparrow} = \frac{P_{\uparrow}}{\pi^2g'^2}; \,\,\,\ \omega'=\frac{\omega}{C_0^2}; \,\,\,\ \nu' = \frac{\nu}{C_0^2}; \,\,\,\ g'=\frac{g}{C_0^2}~,
\label{C8}
\end{eqnarray}
with the change of variable $x=C_0^2 t$.
In Figure \ref{FigC1} we give the numerical plot of this probability. 

\begin{figure}[h!]
	\begin{centering}
		\includegraphics[scale=0.40]{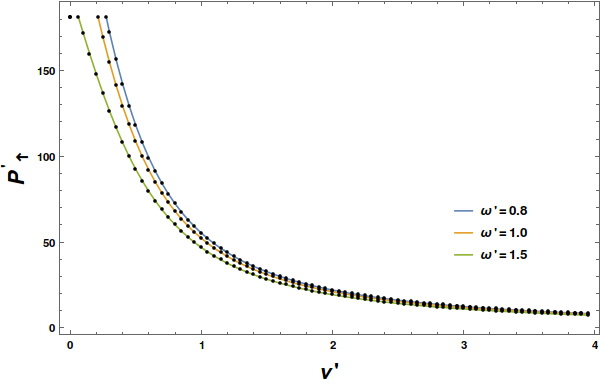}
	\end{centering}
	\caption{Plot of $P'_{\uparrow}$ Vs $\nu'$ for different values of $\omega'$. Here we choose the small parameter $\epsilon=0.01$.}
	\label{FigC2}
\end{figure}

\underline{Matter dominated era}:
Here the transition probability turns out to be
\begin{equation}
P'_{\uparrow} = \frac{1}{\nu'^2}\Big|\int_0^{\infty} \frac{dx}{x} e^{-(-i\omega+\epsilon)x}(e^{6i\nu'x^{1/3}}-1)\Big|^2~,
\label{C9}
\end{equation}
where 
\begin{equation}
P'_{\uparrow} = \frac{9P_{\uparrow}}{4\pi^2 g'^2}; \,\,\,\ \omega'= \frac{\omega}{C_0^{3/2}}; \,\,\ \nu' = \frac{\nu}{C_0^{3/2}}; \,\,\,\ g'=\frac{g}{C_0^{3/2}}~,
\label{C10}
\end{equation}
with the use of the change of variable $x=C_0^{3/2}t$.
The numerical plot is given in Figure \ref{FigC2}.

Note that here also we show $P'_{\uparrow}$ VS $\nu'$ plot because of same reason as mentioned in Section \ref{General}. Nevertheless, the nature of our present plots are similar to Bose-Einstein distribution which implies the particle content in conformal vacuum and it is thermal in nature. In all eras of the Universe, we found that the probability of transition amplitute is non-zero. Hence our present infalling detector along null trajectory in cosmic time will see the conformal vacuum, defined in conformal time coordinates, as full of particles. Like $(1+1)$ dimensional case, here also the probability decreases with the increase of detector's frequency. 
\section{\label{Conclusions}Conclusions and Discussions}
In this work we studied the response function of a two level atomic detector which is traveling freely along the null (light like) trajectory under the background of a non-trivial metric in a particular coordinate system. The spacetimes, we have considered here, are $(a)$ $(1+1)$ dimensional general static black hole and $(b)$ FLRW Universe. For FRW case, all three stages of the Universe, de-Sitter, radiation dominated and matter dominated, were accounted. 

For the black hole case, the detector was moving along the null trajectory in Schwarzschild coordinates. It has been observed that the detector detects particles in the {\it Boulware vacuum}. This has been established both analytically under large detector frequency approximation and also numerically (without any approximation). We noticed that the probability of detection decreases with the increase of detector's frequency $\omega$. This analysis has been further extended to any arbitrary dimensional stationary black hole. In this case we considered the near horizon effective metric of the full spacetime and allowed the detector to move radially within this near horizon region. Again we found the probability of transition is finite, indicating the creation of particle in the Boulware vacuum with respect to this specific observer. In all numerical analysis, the plot between $\nu^2 P(\nu)$ VS $\nu$ is similar to Planck spectrum and the peak decreases with the increase of $\omega$. We also presented a similar analysis in $(1+3)$ dimensional FLRW Universe.       

In this approach we further investigated the three stages of FLRW Universe. We allowed the detector to move along null trajectory in original FLRW coordinates ($t,x$) and massless scalar modes, under study, are corresponding to {\it conformal vacuum}. In all three stages the conformal vacuum appears to be filled with thermal particle with respect to this null infalling observer. The de-Sitter era was analytically solvable and gives perfect black body spectrum. Whereas, radiation and matter dominated eras were investigated numerically. These also showed the particle content in conformal vacuum with respect to our chosen class of observers. Here also the probability decreases with increase of $\omega$. 

It must be mentioned that our present investigation is an extension of the work of Scully et al \cite{Scully:2017utk}, where the authors investigated the issue for the detector moving along timelike geodesic in Schwarzschild spacetime. Here we took  different paths, the null like ones and also discussed the situation for other black holes in arbitrary dimensions. Of course, in this case the analysis is valid only in the near horizon region. We further investigated the three stages of FLRW universe. The particle production in this case has been attempted earlier by different method; but not by detector response method and for this different class of observers (for example see, \cite{Singh:2013bsa,Modak:2018crw,Modak:2018usa}; also see \cite{Modak:2019jbg} for a recent review). In this sense, the present investigation is different from earlier ones and further enlighten the observer dependence of particle production in the case of curved spacetimes.    


\begin{appendix}
\section{\label{App1}Alternative approach: Fourier transformation in momentum space}
\subsection{Black hole}
For our null trajectory (\ref{2.15}), the outgoing massless scaler field mode (\ref{2.20}), with respect to the radially infalling observer, takes the following form:
\begin{equation}
u_\nu (r) = e^{2ir\nu} \Big(\frac{r}{r_H} - 1\Big)^{2ir_H\nu}~.
\label{A1.01}
\end{equation}
Now we are interested to find the corresponding function in the momentum space. This is given by the following Fourier transformation:
\begin{equation}
F(\omega,\nu) = \int_{-\infty}^{\infty} dr e^{i\omega r}u_{\nu}(r)~.
\label{A1.02}
\end{equation}
Now note that the outgoing scalar modes only exists in region outside the horizon and so only the non-vanishing limit of integration of the above integral is $r_H$ to $\infty$. With this, substitution of (\ref{A1.01}) in (\ref{A1.02}) and then use of the change of integration variable $(r/r_H)-1=x$ yield
\begin{equation}
F(\omega,\nu) = r_H e^{i(\omega+2\nu)r_H}\int_0^{\infty}dx e^{ir_H(\omega+2\nu)x} (x)^{2ir_H\nu}~.
\label{A1.03}
\end{equation}
Modulus square of the above expression is our required quantity, we are looking for. With the substitution $r_H(\omega+2\nu)x=z$ one finds
\begin{equation}
\Big|F(\omega,\nu)\Big|^2 = \frac{1}{(\omega+2\nu)^2}\Big|\int_0^{\infty}dz e^{-iz}(z)^{-2ir_H\nu}\Big|^2~.
\label{A1.04}
\end{equation}
Finally using (\ref{2.25}) we obtain
\begin{equation}
\Big|F(\omega,\nu)\Big|^2 = \frac{4\pi r_H\nu}{(\omega+2\nu)^2}\frac{1}{e^{4\pi r_H\nu}-1}~.
\label{A1.05}
\end{equation}
Here again we see that the above expression represents the thermal behaviour of the Boulware modes at temperature $T=1/(4\pi r_H)$ with respect to the radially infalling observer.

\subsection{FLRW Universe}
In the case of FLRW Univese, the conformal mode (\ref{2.41}) for null path (\ref{2.42}) takes the form
\begin{equation}
u_\nu=e^{-2i\nu\eta}~,
\label{A1.06}
\end{equation} 
which in Fourier space of $t$ coordinate is given by the argument of the expression (\ref{2.43}). Therefore the square of the absolute value of this for different stages of the Universe again coincides with the results obtained in Section \ref{FRW}. So the Fourier trick approach also favors the particle production in conformal vacuum with respect to our infalling null observer. 
 

\section{Observer's metric and particle creation through Bogolubov coefficients}
\label{App2}
The actual event of particle production is always perceived by calculating the number operator for the vacuum under study. The detector's click is not always the confirmation of particle production. For example, a uniformly rotating detector in Minkowski spacetime always shows transition; whereas the calculation of number operator in Minkowski vacuum gives vanishing result \cite{Paddy1}. Therefore, it is always necessary to examine the number operator in order to discuss the observer dependent particle production. This is determined by one of the Bogolubov exponent, namely $\beta_{\nu\omega}$ (see chapter $3$  of \cite{Book1}). In this case one needs to find two sets of observers and the corresponding field modes. In our present cases one observer is Boulware for black hole or conformal for FLRW. The other one is moving along the null trajectory. To find the Bogolubov coefficient, first it is necessary to obtain the field modes in two frames. Here in all situations one set of modes are known which are defined in $(t,r^*)$ coordinates for black hole (Eq. (\ref{2.20})) while for FLRW they are in conformal coordinates $(\eta,x)$ (see Eq. (\ref{2.41})). The other set of modes are not known to us at this moment as the metric for the null infalling observer has not been introduced till now. In the below we shall construct the metrics for this observer case by case and find the modes.

\subsection{Schwarzschild black hole}
It must be noted that for the ingoing radial null path (\ref{2.15}), the ingoing null coordinate $v=t_s+r^*$ is constant. Whereas the outgoing null coordinate is given by 
\begin{equation}
u=t_s-r^* = -2r-2r_H\ln((r/r_H) -1) =-2t_s~.
\label{B1.01}
\end{equation} 
So the metric (\ref{2.18}), written in $(u,v)$ coordinates
\begin{equation}
ds^2=-\frac{f(u,v)}{2}(dudv+dvdu)~,
\label{B1.02}
\end{equation}
for the Boulware observer, will be for our detector observer by the following coordinate transformations:
\begin{equation}
dv \rightarrow dv; \,\,\,\ du =-2dt_s=-\frac{2dr}{f}~.
\label{B1.03}
\end{equation}
The last one is obtained by differentiating (\ref{B1.01}). Under this the metric (\ref{B1.02}) takes the following conformally flat form:
\begin{equation}
ds^2 = (drdv+dvdr)~.
\label{B1.04}
\end{equation}
Under the background the outgoing massless scalar mode is (\ref{2.20}); i.e. $u_\nu=(1/\sqrt{2\pi\cdot 2\nu})e^{-i\nu u}$, while that for the above metric is $(1/\sqrt{2\pi\cdot 2\omega})e^{i\omega r}$. $u_\nu$ is normalised for the range $u=-\infty$ to $u=+\infty$; on the other hand $u_\omega$ is normalised for the range $r=\infty$ to $r=r_H$.

Now expand the one mode in terms of other as
\begin{equation}
\frac{1}{\sqrt{\nu}}e^{-i\nu u} = \int_0^{\infty}\frac{d\omega}{\sqrt{\omega}}\Big(\alpha_{\omega\nu}e^{ir\omega}-\beta^*_{\omega\nu}e^{-ir\omega}\Big)~,
\label{B1.06}
\end{equation}
where the Bogolubov coefficients are determined as
\begin{eqnarray}
&&\alpha_{\omega\nu} = \frac{1}{2\pi}\sqrt{\frac{\omega}{\nu}}\int_{r_H}^\infty dr e^{-i\nu u - ir\omega}~;
\nonumber
\\
&&\beta^*_{\omega\nu} = -\frac{1}{2\pi}\sqrt{\frac{\omega}{\nu}}\int_{r_H}^\infty dr e^{-i\nu u + ir\omega}~.
\label{B1.07}
\end{eqnarray}
Following the earlier steps we find
\begin{eqnarray}
&&\Big|\alpha_{\omega\nu}\Big|^2 = \frac{\omega r_H}{\pi(\omega-2\nu)^2}\frac{1}{1-e^{-4\pi r_H\nu}}~;
\nonumber
\\
&&\Big|\beta_{\omega\nu}\Big|^2 = \frac{\omega r_H}{\pi(\omega+2\nu)^2}\frac{1}{e^{4\pi r_H\nu}-1}~.
\label{B1.08}
\end{eqnarray}
To get the above results we considered $\omega>2\nu$.
Here we obtain the non-vanishing value of $|\beta_{\omega\nu}|^2$ whose structure is similar to Planck distribution with temperature $T=1/(4\pi r_H)$. This implies the actual particle production in the Boulware vacuum as seen by our particular observer.

\subsection{FLRW Universe}
For FLRW case, one set of observers (conformal) is using $u=\eta-x$ and $v=\eta+x$ coordinates and the metric is
\begin{equation}
ds^2 = -\frac{a^2}{2}(dudv+dvdu)~.
\label{B1.09}
\end{equation} 
The infalling observer's coordinates are related to those for conformal one as follows
\begin{equation}
dv\rightarrow dv; \,\,\,\ du = d\eta-dx=2\eta=\frac{2dt}{a}~,
\label{B1.010}
\end{equation} 
and correspondingly the metric is 
\begin{equation}
ds^2=-a(dtdv+dvdt)~.
\label{B1.11}
\end{equation}
The relevant outgoing modes are $u_\nu = (1/\sqrt{2\pi\cdot 2\nu})e^{-i\nu u}$ and $u_\omega=(1/\sqrt{2\pi\cdot 2\omega})e^{-i\omega t}$. In this case the Bogolubov coefficients are given by
\begin{eqnarray}
&&\alpha_{\omega\nu} = \frac{1}{2\pi}\sqrt{\frac{\omega}{\nu}}\int dt e^{-i\nu u + i\omega t}~;
\nonumber
\\
&&\beta^*_{\omega\nu} = -\frac{1}{2\pi}\sqrt{\frac{\omega}{\nu}}\int dt e^{-i\nu u - i\omega t}~.
\label{B1.12}
\end{eqnarray} 
These are the same integration, which were identified as the detector's response function $P_{\uparrow}$ in Section \ref{FRW}, particularly $|\beta_{\omega\nu}|^2$. Clearly, these are non-vanishing and the infalling observer will see particles in the conformal vacuum.

\end{appendix}


\end{document}